\font\bbb=msbm10                                                   

\def\AnM{{\sl Ann.\ Math.}}

\def\Di{{\sl Dialectica\/}}

\def\DMJDMV{{\sl Doc.\ Math.\ J. DMV}}

\def\JMM{{\sl J. Math.\ Mech.}}

\def\JPA{{\sl J. Phys.\ A:  Math.\ Gen.}}

\def\Na{{\sl Nature\/}}

\def\Ph{{\sl Physics\/}}

\def\PJM{{\sl Pacific J. Math.}}
\def\PLA{{\sl Phys.\ Lett.\ A\/}}

\def\PNAWA{{\sl Proc.\ Nederl.\ Akad.\ Wetensch., Ser.\ A}}

\def\PR{{\sl Phys.\ Rev.}}
\def\PRA{{\sl Phys.\ Rev.\ A\/}}

\def\PRD{{\sl Phys.\ Rev.\ D\/}}

\def\PRL{{\sl Phys.\ Rev.\ Lett.}}

\def\PS{{\sl Philos.\ Sci.}}

\def\PhT{{\sl Phys.\ Today\/}}

\def\RMP{{\sl Rev.\ Mod.\ Phys.}}

\def\SIAMJC{{\sl SIAM J. Comput.}}

\def\pitowsky{\hbox{I. Pitowsky}}

\def\hfb{\hfil\break}

\catcode`@=11
\newskip\ttglue

   \font\ninerm=cmr9    \font\eightrm=cmr8   \font\sixrm=cmr6
  \font\ninebf=cmbx9   \font\eightbf=cmbx8  \font\sixbf=cmbx6
  \font\nineit=cmti9   \font\eightit=cmti8  
  \font\ninesl=cmsl9   \font\eightsl=cmsl8  
  \font\ninemi=cmmi9   \font\eightmi=cmmi8  \font\sixmi=cmmi6

\font\bigten=cmr10 scaled\magstep2 

\def\ninepoint{\def\rm{\fam0\ninerm}%
  \textfont0=\ninerm \scriptfont0=\sixrm
  \textfont1=\ninemi \scriptfont1=\sixmi
  \textfont\itfam=\nineit  \def\it{\fam\itfam\nineit}%
  \textfont\slfam=\ninesl  \def\sl{\fam\slfam\ninesl}%
  \textfont\bffam=\ninebf  \scriptfont\bffam=\sixbf
    \def\bf{\fam\bffam\ninebf}%
  \tt \ttglue=.5em plus.25em minus.15em
  \normalbaselineskip=11pt
  \setbox\strutbox=\hbox{\vrule height8pt depth3pt width0pt}%
  \normalbaselines\rm}

\def\eightpoint{\def\rm{\fam0\eightrm}%
  \textfont0=\eightrm \scriptfont0=\sixrm
  \textfont1=\eightmi \scriptfont1=\sixmi
  \textfont\itfam=\eightit  \def\it{\fam\itfam\eightit}%
  \textfont\slfam=\eightsl  \def\sl{\fam\slfam\eightsl}%
  \textfont\bffam=\eightbf  \scriptfont\bffam=\sixbf
    \def\bf{\fam\bffam\eightbf}%
  \tt \ttglue=.5em plus.25em minus.15em
  \normalbaselineskip=9pt
  \setbox\strutbox=\hbox{\vrule height7pt depth2pt width0pt}%
  \normalbaselines\rm}

\def\sfootnote#1{\edef\@sf{\spacefactor\the\spacefactor}#1\@sf
      \insert\footins\bgroup\eightpoint
      \interlinepenalty100 \let\par=\endgraf
        \leftskip=0pt \rightskip=0pt
        \splittopskip=10pt plus 1pt minus 1pt \floatingpenalty=20000
        \parskip=0pt\smallskip\item{#1}\bgroup\strut\aftergroup\@foot\let\next}
\skip\footins=12pt plus 2pt minus 2pt
\dimen\footins=30pc

\def\ie{{\it i.e.}}

\def\Proof{{\sl Proof}}
\def\Lemma{L{\eightpoint EMMA}}

\def\endproof{\vrule height6pt width6pt depth0pt}

\def\Q{\hbox{\bbb Q}}
\def\R{\hbox{\bbb R}}

\def\Horodeccy{1}
\def\teleportation{2}
\def\downconversion{3}
\def\EPRB{4}
\def\Bellepr{5}
\def\CHSH{6}
\def\NMR{7}
\def\earlyEPRB{8}
\def\KokBraunstein{9}
\def\Bellgleason{10}
\def\Gleason{11}
\def\KochenSpecker{12}
\def\GHZM{13}
\def\communication{14}
\def\Specker{15}
\def\deBruijnErdos{16}
\def\Peres{17}
\def\PeresKC{18}
\def\BirkhoffvonNeumann{19}
\def\Pitowskycoloring{20}
\def\Pitowskyconj{21}
\def\separability{22}
\def\Jauch{23}
\def\GodsilZaks{24}
\def\HalesStraus{25}
\def\Kent{26}
\def\Schonhage{27}
\def\Freedman{28}
\def\Shor{29}
\def\CliftonKent{30}

\magnification=1200
\input epsf.tex

\dimen0=\hsize \divide\dimen0 by 13 \dimendef\chasm=0
\dimen1=\hsize \advance\dimen1 by -\chasm \dimendef\usewidth=1
\dimen2=\usewidth \divide\dimen2 by 2 \dimendef\halfwidth=2
\dimen3=\usewidth \divide\dimen3 by 3 \dimendef\thirdwidth=3
\dimen4=\hsize \advance\dimen4 by -\halfwidth \dimendef\secondstart=4
\dimen5=\halfwidth \advance\dimen5 by -10pt \dimendef\indenthalfwidth=5
\dimen6=\thirdwidth \multiply\dimen6 by 2 \dimendef\twothirdswidth=6
\dimen7=\twothirdswidth \divide\dimen7 by 4 \dimendef\qttw=7
\dimen8=\qttw \divide\dimen8 by 4 \dimendef\qqttw=8
\dimen9=\qqttw \divide\dimen9 by 4 \dimendef\qqqttw=9

\parskip=0pt\parindent=0pt

\line{\hfil 25 March 1999}
\line{\hfil {\it revised\/} 19 September 1999}
\line{\hfil quant-ph/9905080}
\vfill
\centerline{\bf\bigten FINITE PRECISION MEASUREMENT NULLIFIES}
\bigskip
\centerline{\bf\bigten THE KOCHEN-SPECKER THEOREM}
\bigskip\bigskip
\centerline{\bf David A. Meyer}
\bigskip 
\centerline{\sl Project in Geometry and Physics}
\centerline{\sl Department of Mathematics}
\centerline{\sl University of California/San Diego}
\centerline{\sl La Jolla, CA 92093-0112}
\centerline{dmeyer@chonji.ucsd.edu}
\smallskip
\centerline{\sl and Institute for Physical Sciences} 
\centerline{\sl Los Alamos, NM}
\vfill
\centerline{ABSTRACT}
\bigskip
\noindent Only finite precision measurements are experimentally 
reasonable, and they cannot distinguish a dense subset from its 
closure.  We show that the rational vectors, which are dense in $S^2$, 
can be colored so that the contradiction with hidden variable theories 
provided by Kochen-Specker constructions does not obtain.  Thus, in 
contrast to violation of the Bell inequalities, no 
quantum-over-classical advantage for information processing can be 
derived from the Kochen-Specker theorem alone.

\bigskip\bigskip
\noindent 1999 Physics and Astronomy Classification Scheme:
                   03.65.Bz, 
                   03.67.Hk, 
                   03.67.Lx. 

\noindent American Mathematical Society Subject Classification:
                   81P10,    
                   03G12,    
                   68Q15.    

\smallskip
\global\setbox1=\hbox{Key Words:\enspace}
\parindent=\wd1
\item{Key Words:}  quantum computation, Kochen-Specker theorem, hidden
                   variable theories, quantum communication 
                   complexity.
 
\vfill\eject

\headline{\ninepoint\it Finite precision nullifies the K-S theorem  
                                                 \hfil David A. Meyer}
\parskip=10pt
\parindent=20pt

Recent theoretical and experimental work on quantum computation and 
quantum information theory has inspired renewed interest in 
fundamental results of quantum mechanics.  The Horodeccy have shown, 
for example, that a spin-${1\over2}$ state can be teleported with
greater than classical fidelity using any mixed two spin-${1\over2}$
state which violates some generalized Bell-CHSH inequality 
[\Horodeccy].  Quantum teleportation was first demonstrated 
experimentally in quantum optics systems [\teleportation]; the 
parametric down-conversion techniques crucial for these experiments 
have also been used to verify violation of Bell's inequality
directly [\downconversion].  Although the Bell-CHSH inequalities were
originally derived in the context of EPR-B experiments [\EPRB] and
(local) hidden variable theories [\Bellepr,\CHSH], the present concern 
is with the differences in information processing capability between 
classical and quantum systems.%
\sfootnote{$^*$}{See, for example, the recent discussion of 
                 separability in NMR experiments [\NMR]; at issue is
                 whether these perform or merely simulate quantum 
                 computations.}
%

Analyses of EPR-B experiments from the very first [\earlyEPRB] have 
been concerned with limitations in, for example, detector efficiency 
[\CHSH]:  The observed violations of Bell-CHSH inequalities are 
consequently reduced; so, too, is teleportation fidelity 
[\teleportation,\KokBraunstein].

Logically, if not entirely chronologically prior contradictions with
(noncontextual) hidden variable theories were derived by Bell 
[\Bellgleason] from a theorem of Gleason [\Gleason] and by Kochen and
Specker [\KochenSpecker].  The GHZ-Mermin three spin-${1\over2}$ state 
exhibits a similar incompatibility with (noncontextual) hidden 
variable theories [\GHZM] and reduces the communication complexity of 
certain problems [\communication].  While no quantum improvement in 
information processing power has yet been derived directly from the 
Kochen-Specker theorem, it is natural to ask for the consequences of 
experimental limitations on measurement in this setting.

The Kochen-Specker theorem concerns the results of a (counterfactual) 
set of measurements on a quantum system described by a vector in a 
three dimensional Hilbert space.  Kochen and Specker consider, for 
example, measurement of the squares of the three angular momentum 
components of a spin-1 state [\KochenSpecker].  The corresponding 
operators commute and can be measured simultaneously, providing one 
`yes' and two `no's to the three questions, ``Does the spin component 
along $\hat a$, $\hat b$, $\hat a\times\hat b$ vanish?'' for any 
$\hat a\perp\hat b \in S^2$, the unit sphere in $\R^3$.  Specker 
[\Specker] and Bell [\Bellgleason] observed that Gleason's theorem 
[\Gleason] implies that there can be no assignment of `yes's and `no's 
to the vectors of $S^2$ consistent with this requirement:
$$
\hbox{each triad is `colored' with one `yes' and two `no's}   \eqno(1)
$$
(where {\sl triad\/} means three mutually orthogonal vectors) and 
concluded that there could be no theory with hidden variables assigned 
independently of the measurement context.

A compactness argument [\deBruijnErdos] implies that there must be a 
{\sl finite\/} set of triads for which there is no coloring satisfying 
(1).  Kochen and Specker gave the first explicit construction of a 
such a finite set [\KochenSpecker].  Their construction requires 117 
vectors; subsequently examples with 33 [\Peres] and 31 [\PeresKC] 
vectors in $S^2$ have been constructed.

For our present purposes the details of these constructions are 
unimportant; all that matters is that the vectors forming the set of
triads are precisely specified.  But, as Birkhoff and von~Neumann 
remark in their seminal study of the lattice of projections in quantum 
mechanics, ``it seems best to assume that it is the 
{\sl Lebesgue-measurable\/} subsets $\ldots$ which correspond to 
experimental propositions, two subsets being identified, if their 
difference has {\sl Lebesgue-measure\/} 0.'' 
[\BirkhoffvonNeumann, p.825]\ \ That is, no experimental arrangement 
could be aligned to measure spin projections along coordinate axes 
specified with more than finite precision.  The triads of a 
Kochen-Specker construction should therefore be constrained only to 
lie within some (small) neighborhoods of their ideal positions.  This 
is sufficient to nullify the Kochen-Specker theorem because, as we 
will show presently, there is a coloring of the vectors of a set of 
triads, dense in the space of triads, which respects (1).  More 
complicated colorings satisfying (1) `almost everywhere' have been
constructed by Pitowsky using the axiom of choice and the continuum
hypothesis [\Pitowskycoloring]; our results here support a conjecture
of his that many dense subsets---in particular, the rational 
vectors---have colorings which satisfy (1) [\Pitowskyconj], but we 
will need no more than constructive set theory.

The finite constructions violating (1) provide the clue we use:  in 
each case the components of some of the vectors forming triads are
irrational numbers.  So let us consider only the vectors with rational
components, $S^2 \cap \Q^3$.  This is a familiar subset:  the usual
requirement of separability%
\sfootnote{$^*$}{To avoid possible confusion, we remark that this is a
                 distinct concept from that of separability of density
                 matrices [\separability].}
for Hilbert space and for the lattice of measurement propositions 
depends upon such a countable dense subset [\Jauch]; the fact that it
is dense means that it is indistinguishable from its closure by finite
precision measurements.  As Jauch puts it, while the rationals must 
already be defined with infinite precision, completing them to include 
the irrationals requires that ``we transcend the proximably observable 
facts and $\ldots$ introduce {\sl ideal elements\/} into the 
description of physical systems.''~[\Jauch, p.75]\ \ Surely the 
meaning of quantum mechanics should not rest upon such 
non-experimental entities.  But, at least in the three dimensional 
arena for the Kochen-Specker theorem it does, as we will be able to 
conclude from the following three lemmas:

\noindent\Lemma\ 1.  {\sl The rational vectors $S^2 \cap \Q^3$ can be
colored to satisfy\/} ({\sl 1\/}).

\noindent\Proof.  This is an immediate consequence of a result of 
Godsil and Zaks [\GodsilZaks] which is in turn based upon a theorem of
Hales and Straus [\HalesStraus].  It suffices here to give an explicit 
coloring using their results.  The rational projective plane $\Q P^2$ 
consists of triples of integers $(x,y,z)$ with no common factor other 
than 1 (every integer is taken to divide 0).  Since at least one of 
$x$, $y$ and $z$ must therefore be odd, and since odd (even) numbers 
square to 1 (0) modulo 4, exactly one must be odd if $x^2 + y^2 + z^2$ 
is to be a square.  In this case, and only in this case, 
$(x,y,z) \in \Q P^2$ can be identified as a vector in $S^2 \cap \Q^3$.  
Consider a triad of such vectors.  For any two, $(x,y,z)$ and 
$(x',y',z')$, $x'x + y'y + z'z = 0$ implies that they must differ in 
which component is odd.  Thus exactly one vector of any triad has an 
odd $z$ component.  Color this one `yes' and the other two `no'.  This 
defines a {\sl $z$-parity\/} coloring of $S^2 \cap \Q^3$ satisfying 
(1).                                                   \hfill\endproof

The rational vectors are dense in $S^2$ since $\Q^2$ is dense in 
$\R^2$ and rational vectors in $S^2$ map bijectively to rational 
points in the affine plane---since stereographic projection is a 
birational map.  Furthermore:

\noindent\Lemma\ 2.  {\sl The rational vectors $z$-parity colored 
`yes' are dense in $S^2$.}

\noindent\Proof.  Again we follow the argument of Godsil and Zaks 
[\GodsilZaks]:  Rotation by angle $\alpha = \arccos{3\over5}$ about
the $x$-axis takes each rational vector $(0,y,z)$ with odd $z$ (\ie, 
colored `yes') to another rational vector colored `yes'.  Since 
$\alpha$ is not a rational multiple of $\pi$, iterated rotation takes
$(0,0,1)$ to a dense set of vectors in the $x = 0$ great circle of 
$S^2$.  Similarly, iterated rotation by angle $\alpha$ around the 
$z$-axis takes this set of vectors dense in $S^1$ to a set of vectors 
dense in $S^2$, each of which is colored `yes' since it has odd 
$z$-component.                                         \hfill\endproof

Repeating this argument with $(x,y,z)$ permuted to $(y,z,x)$ shows 
that the rational vectors $z$-parity colored `no' are also dense in 
$S^2$.

\noindent\Lemma\ 3.  {\sl The rational triads are dense in the space
of triads.}

\noindent\Proof.  By the proof of Lemma~2, for any $\epsilon > 0$,
within a ${1\over2}\epsilon$-neighborhood of a specified vector 
$\hat a$ of a triad, $\hat a$, $\hat b$, $\hat a \times \hat b$, there 
is a rational vector $\hat u$ to which $(0,0,1)$ is mapped by an 
SO$(3,\Q)$ rotation.  This rotation maps the rational vectors 
$(x,y,0)$ on the the equator to the rational vectors in a great circle 
passing through the ${1\over2}\epsilon$-neighborhoods of $\hat b$ and 
$\hat a \times \hat b$.  Since the rational points are dense in the 
equator (also a consequence of the proof of Lemma~2) there is a 
rational vector $\hat v \perp \hat u$ in the 
${1\over2}\epsilon$-neighborhood of $\hat b$, and thus 
$\hat u \times \hat v$ is a rational vector in the 
$\epsilon$-neighborhood of $\hat a \times \hat b$.     \hfill\endproof

Suppose we measure some triad in a three dimensional Kochen-Specker
construction.  By Lemma~3 the unavoidable finite precision of this 
measurement cannot distinguish it from the (many) rational triads
within some neighborhood of the intended triad.  By Lemmas~1 and 2 the 
results of a (counterfactual) set of such measurements cannot conflict
with (1), and so cannot rule out a noncontextual hidden variable
theory {\sl defined over the rationals}.  Thus finite precision 
measurement nullifies the Kochen-Specker theorem.  The $z$-parity 
coloring of $S^2 \cap \Q^3$ shows that arguments such as Bell's 
[\Bellgleason], based on Gleason's theorem [\Gleason] in three 
dimensions, also fail when the finite precision of measurement is 
taken into account.%
\sfootnote{$^*$}{Kent has recently generalized the results of this 
                 paper to show that similar constructions produce 
                 `colorings' of dense subsets satisfying the analogue 
                 of (1) in all higher dimensional real or complex 
                 Hilbert spaces as well.~[\Kent]}
Although our explicit construction involves the rational vectors, we 
emphasize that they are incidental to the interpretation of this 
result.  {\sl Any\/} dense subset is indistinguishable by finite
precision measurement from its completion, so any colorable dense
subset would do equally well.  Our results, together with Pitowsky's
earlier [\Pitowskycoloring] and Kent's subsequent [\Kent] 
constructions indicate that there are many such subsets.

We conclude by remarking that while one might object that since the
counterfactual measurements specified by a Kochen-Specker construction
are not (simultaneously) experimentally realizable, it is unreasonable
to impose the experimental limitation of finite precision on such a
theoretical edifice.  But theoretical analyses of the power of 
algorithms must address the possibility that it resides in infinite 
precision specification of the computational states or the operations 
on them.  Sch\"onhage showed, for example, that classical computation 
with infinite precision real numbers would solve NP-complete problems 
efficiently [\Schonhage].  And, as Freedman has emphasized, even 
classical statistical mechanics models would solve \#P-hard problems 
were infinite precision measurement possible [\Freedman].  The promise 
of {\sl quantum\/} computation, in contrast, is efficient 
algorithms---which require only poly(log) number of bits 
precision---for problems not known to have polynomial time classical 
solutions [\Shor].  Thus, despite the relation noted earlier with the 
GHZ-Mermin state which can reduce communication complexity, the 
elementary argument presented here shows that given the finite 
precision of any expermental measurement, the Kochen-Specker theorem
alone cannot separate quantum from classical information processing in
three dimensional Hilbert space.  We have not, of course, constructed
even a static (much less a dynamic) hidden variable theory for a 
spin-1 particle, so we have not proved that no separation result is
possible---only that the Kochen-Specker theorem does not imply one, as
we might have expected.  Our results, and Pitowsky's deterministic
model [\Pitowskycoloring], however, make it seem unlikely that any 
separation exists.%
\sfootnote{$^*$}{Confirming this conjecture, Clifton and Kent have 
recently extended the results of this paper and [\Kent] to construct a
noncontextual hidden variable model for finite precision measurements
in any finite dimensional Hilbert space.~[\CliftonKent]}

\medskip
\noindent{\bf Acknowledgements}

\noindent I thank Peter Doyle, Michael Freedman, Chris Fuchs, Asher 
Peres, Jeff Rabin and especially Rafael Sorkin for useful discussions, 
Chris Godsil and Joseph Zaks for providing reference [\GodsilZaks], 
and Philippe Eberhard for suggestions which improved the exposition.  
This work was partially supported by ARO grant DAAG55-98-1-0376.

\bigskip
\global\setbox1=\hbox{[00]\enspace}
\parindent=\wd1

\noindent{\bf References}
\vskip10pt

\parskip=0pt
\item{[\Horodeccy]}
R. Horodecki, M. Horodecki and P. Horodecki,
``Teleportation, Bell's inequalities and inseparability'',
\PLA\ {\bf 222} (1996) 21--25.

\item{[\teleportation]}
D. Bouwmeester, J.-W. Pan, K. Mattle, M. Eibl, H. Weinfurter and 
A. Zeilinger, 
``Experimental quantum teleportation'', 
\Na\ {\bf 390} (1997) 575--579;\hfb
D. Boschi, S. Branca, F. De Martini, L. Hardy and S. Popescu,
``Experimental realization of teleporting an unknown pure quantum 
  state via dual classical and Einstein-Podolsky-Rosen channels'',
\PRL\ {\bf 80} (1998) 1121--1125.

\item{[\downconversion]}
Z. Y. Ou and L. Mandel,
``Violation of Bell's inequality and classical probability in a 
  two-photon correlation experiment'',
\PRL\ {\bf 61} (1988) 50--53;\hfb
Y. H. Shih and C. O. Alley,
``New type of Einstein-Podolsky-Rosen-Bohm experiment using pairs of
  light quanta produced by optical parametric down conversion'',
\PRL\ {\bf 61} (1988) 2921--2924;\hfb
T. E. Kreiss, Y. H. Shih, A. V. Sergienko and C. O. Alley,
``Einstein-Podolsky-Rosen-Bohm experiment using pairs of light quanta
  produced by type-II parametric down-conversion'',
\PRL\ {\bf 71} (1993) 3893--3897.

\item{[\EPRB]}
A. Einstein, B. Podolsky and N. Rosen,
``Can quantum-mechanical description of physical reality be 
considered complete?'',
\PR\ {\bf 47} (1935) 777--780;\hfb
D. Bohm,
{\sl Quantum Theory\/}
(New York:  Prentice-Hall 1951).

\item{[\Bellepr]}
J. S. Bell,
``On the Einstein-Podolsky-Rosen paradox'',
\Ph\ {\bf 1} (1964) 195--200.

\item{[\CHSH]}
J. F. Clauser, M. A. Horne, A. Shimony and R. A. Holt,
``Proposed experiment to test local hidden-variable theories'',
\PRL\ {\bf 23} (1969) 880--884.

\item{[\NMR]}
S. L. Braunstein, C. M. Caves, R. Jozsa, N. Linden, S. Popescu and
R. Schack,
``Separability of very noisy mixed states and implications for NMR
  quantum computing'',
\PRL\ {\bf 83} (1999) 1054--1057;\hfb
R. Laflamme,
in {\sl Quick Reviews in Quantum Computation and Information},\hfb
{\tt http://quantum-computing.lanl.gov/qcreviews/qc/},
15 January 1999;\hfb
R. Schack and C. M. Caves,
``Classical model for bulk-ensemble NMR quantum computation'',
quant-ph/9903101.

\item{[\earlyEPRB]}
C. S. Wu and I. Shaknov,
``The angular correlation of scattered annihilation radiation'',
\PR\ {\bf 77} (1950) 136;\hfb
C. A. Kocher and E. D. Commins,
``Polarization correlations of photons emitted in an atomic 
  cascade'',
\PRL\ {\bf 18} (1967) 575--577;\hfb
A. Aspect, P. Grangier and G. Roger,
``Experimental tests of realistic local theories via Bell's
  theorem'',
\PRL\ {\bf 47} (1981) 460--463.

\item{[\KokBraunstein]}
P. Kok and S. L. Braunstein,
``On quantum teleportation using parametric down-conversion'',
quant-ph/9903074.

\item{[\Bellgleason]}
J. S. Bell,
``On the problem of hidden variables in quantum mechanics'',
\RMP\ {\bf 38} (1966) 447--452.

\item{[\Gleason]}
A. M. Gleason,
``Measures on the closed subspaces of a Hilbert space'',
\JMM\ {\bf 6} (1957) 885--893.

\item{[\KochenSpecker]}
S. Kochen and E. P. Specker,
``The problem of hidden variables in quantum mechanics'',
\JMM\ {\bf 17} (1967) 59--87.

\item{[\GHZM]}
D. M. Greenberger, M. A. Horne and A. Zeilinger,
``Going beyond Bell's theorem'',
in M. Kafatos, ed.,
{\sl Bell's Theorem, Quantum Theory and Conceptions of the
  Universe\/}
(Boston:  Kluwer 1989) 69--72;\hfb
N. D. Mermin,
``What's wrong with these elements of reality'',
\PhT\ {\bf 43} (June 1990) 9--11.

\item{[\communication]}
R. Cleve and H. Buhrman,
``Substituting quantum entanglement for communication'',
\PRA\ {\bf 56} (1997) 1201--1204;\hfb
H. Buhrman, R. Cleve and W. van Dam,
``Quantum entanglement and communication complexity'',
quant-ph/9705033.

\item{[\Specker]}
E. Specker,
``{\it Die Logik nicht gleichzeitig entscheidbarer Aussagen}'',
\Di\ {\bf 14} (1960) 239--246.

\item{[\deBruijnErdos]}
N. G. de Bruijn and P. Erd\"os,
``A color problem for infinite graphs and a problem in the theory
  of relations'',
\PNAWA\ {\bf 54} (1951) 371--373.

\item{[\Peres]}
A. Peres,
``Two simple proofs of the Kochen-Specker theorem'',
\JPA\ {\bf 24} (1991) L175--L178.

\item{[\PeresKC]}
A. Peres,
{\sl Quantum Theory:  Concepts and Methods\/}
(Boston:  Kluwer 1995) p.197.

\item{[\BirkhoffvonNeumann]}
G. Birkhoff and J. von~Neumann,
``The logic of quantum mechanics'',
\AnM\ {\bf 37} (1936) 823--843.

\item{[\Pitowskycoloring]}
\pitowsky,
``Deterministic model of spin and statistics'',
\PRD\ {\bf 27} (1983) 2316--2326;\hfb
\pitowsky,
``Quantum mechanics and value definiteness'',
\PS\ {\bf 52} (1985) 154--156.

\item{[\Pitowskyconj]}
\pitowsky,
email responding to a question from C. A. Fuchs (1998).

\item{[\separability]}
A. Peres,
``Separability criterion for density matrices'',
\PRL\ {\bf 77} (1996) 1413--1415;\hfb
M. Horodecki, P. Horodecki and R. Horodecki,
``Separability of mixed states:  necessary and sufficient 
  conditions'',
\PLA\ {\bf 223} (1996) 1--8.

\item{[\Jauch]}
J. M. Jauch,
{\sl Foundations of Quantum Mechanics\/}
(Menlo Park, CA:  Addison-Wesley 1968).

\item{[\GodsilZaks]}
C. D. Godsil and J. Zaks,
``Colouring the sphere'',
University of Waterloo research report CORR 88-12 (1988).

\item{[\HalesStraus]}
A. W. Hales and E. G. Straus,
``Projective colorings'',
\PJM\ {\bf 99} (1982) 31--43.

\item{[\Kent]}
A. Kent,
``Non-contextual hidden variables and physical measurements'',
quant-ph/ 9906006.

\item{[\Schonhage]}
A. Sch\"onhage,
``On the power of random access machines'',
{\sl Automata, Languages and Programming\/}
(Sixth Colloquium, Graz, 1979),
Lecture Notes in Computer Science, Vol.~71 
(New York:  Springer 1979) 520--529.

\item{[\Freedman]}
M. H. Freedman,
``Topological views on computational complexity'',
{\sl Proceedings of the International Congress of Mathematicians}, 
Vol. II (Berlin, 1998), 
\DMJDMV, Extra Vol.\ ICM II (1998) 453--464.

\item{[\Shor]}
P. W. Shor,
``Polynomial-time algorithms for prime factorization and discrete 
  logarithms on a quantum computer'',
\SIAMJC\ {\bf 26} (1997) 1484--1509.

\item{[\CliftonKent]}
R. Clifton and A. Kent,
``Simulating quantum mechanics by non-contextual hidden variables'',
quant-ph/9908031.

\bye